\documentclass[aps,onecolumn,preprint,superscriptaddress,nofootinbib,12pt]{revtex4-1}
\usepackage{times}
\usepackage{zref}
\usepackage[dvipsnames]{xcolor}
\usepackage{slashed}
\usepackage{amsfonts}
\usepackage{amssymb}
\usepackage{amsmath}
\usepackage{amsopn}
\usepackage{xspace}
\usepackage{bm}
\usepackage{MnSymbol}
\usepackage{array}
\usepackage{epsfig}
\usepackage{mathrsfs}
\usepackage{graphics}
\usepackage{epsfig}
\usepackage{natbib,hyperref}
\definecolor{blue}{HTML}{4169E1}
\definecolor{red}{HTML}{DC143C}
\definecolor{green}{HTML}{2E8B57}
\definecolor{black}{HTML}{000000}
\definecolor{g1}{HTML}{A9A9A9}
\definecolor{g2}{HTML}{696969}
\definecolor{g3}{HTML}{7F7F7F}
\definecolor{g4}{HTML}{D3D3D3}
\newcommand{\mpis}{$m_\pi=137~${\small MeV}}
\newcommand{\mpim}{$m_\pi=450~${\small MeV}}
\newcommand{\mpil}{$m_\pi=806~${\small MeV}}
\newcommand{\muh}{\mu_{^3\text{\scriptsize He}}}
\newcommand{\mut}{\mu_{^3\text{\scriptsize H}}}
\newcommand{\mud}{\mu_\text{\scriptsize D}}
\newcommand{\pode}{\beta_{\text{\scriptsize D}}}
\newcommand{\poh}{\beta_{^3\text{\scriptsize He}}}
\newcommand{\pot}{\beta_{^3\text{\scriptsize H}}}

\newcommand{\eg}{\textit{e.g.}\;}
\newcommand{\ie}{\textit{i.e.}\;}
\newcommand{\be}{\begin{equation}}
\newcommand{\ee}{\end{equation}}

\newcommand{\ber}{\begin{eqnarray}}
\newcommand{\eer}{\end{eqnarray}}

\newcommand{\bs}[1]{\ensuremath{\boldsymbol{#1}}}
\newcommand{\bea}{\begin{eqnarray}}
\newcommand{\eea}{\end{eqnarray}}
\newcommand{\beq}{\begin{align}}
\newcommand{\eeq}{\end{align}}

\newcommand{\bt}{B_{^{3}\text{H}}}
\newcommand{\bh}{B_{^{3}\text{He}}}
\newcommand{\bd}{B_\text{D}}
\newcommand{\rgm}{$\mathbb{R}$GM}

\newcommand{\nopi}{\pi\hspace{-6pt}/}
\newcommand{\ve}[1]{\ensuremath{\boldsymbol{#1}}}

\newcommand{\sgmvec}{\ensuremath{\boldsymbol{\sigma}}}
\newcommand{\tauvec}{\ensuremath{\boldsymbol{\tau}}}

\newcommand{\bra}{\langle~}
\newcommand{\ket}{\rangle}
\newcommand{\eftnopi}{\mbox{EFT($\slashed{\pi}$)}}

\begin{document}

\title{Electromagnetic characteristics\\ of $A\leq3$ physical and lattice nuclei}

\author{Johannes Kirscher}
\affiliation{Racah Institute of Physics, The Hebrew University, 91904, Jerusalem, Israel}
\affiliation{Department of Physics, The City College of New York, New York, NY 10031, USA}
\author{Ehoud Pazy}
\affiliation{Department of Physics, NRCN, P.O.B 9001,Beer Sheva 84190, Israel}
\author{Jonathan Drachman}
\affiliation{Racah Institute of Physics, The Hebrew University, 91904, Jerusalem, Israel}
\author{Nir Barnea}
\affiliation{Racah Institute of Physics, The Hebrew University, 91904, Jerusalem, Israel}

\begin{abstract}
We analyze the
quark-mass dependence of electromagnetic properties of
two and three-nucleon states.
To that end, we apply the pionless effective field theory to experimental data and
numerical lattice calculations which simulate QCD at pion masses
of 450~MeV and 806~MeV.

At the physical pion mass, we postdict the magnetic moment of
helium-3, \mbox{$\muh=-2.13~$nNM}, and the magnetic polarizability
of deuterium, \mbox{$\pode=7.33~10^{-2}~$fm$^3$}.
Magnetic polarizabilities of helium-3, \mbox{$\poh=9.7~10^{-4}~$fm$^3$}, and
the triton, \mbox{$\pot=8.2~10^{-4}~$fm$^3$}, are predictions.

Postdictions of the effective theory for the magnetic moments are found consistent with
QCD simulations at 806~MeV pion mass
while our EFT result \mbox{$\pode=2.92~10^{-2}~$fm$^3$} was not extracted from
the lattice.
The deuteron would thus be relatively pliable compared to a three-nucleon state
for which we postdict \mbox{$\pot=3.9~10^{-5}~$fm$^3$}.
At $m_\pi=450~$MeV, the magnetic moment of the triton is
predicted, \mbox{$\muh=-2.15(5)~$nNM}, based on a conjecture of its binding energy, 
\mbox{$\bt\cong 30$~MeV}.

For all three pion masses, we compare the point-charge radii of the two and
three-nucleon bound states.
The sensitivity of the electromagnetic properties
to the Coulomb interaction between protons is studied 
in anticipation of lattice
calculations with dynamical QED. 

%Our analysis is leading order in the fine structure constant,
%and we estimate the uncertainty by variation of a cutoff regulator.

\end{abstract}
\bigskip
\maketitle
%\tableofcontents
% --
%=============================================================================
\section{Overture} \label{Sec:overt}
%=============================================================================
%The model for nuclei being composites of elementary neutrons and
%protons was found consistent with the Standard Model of particle
%physics through numerical lattice solutions of quantum chromodynamics (LQCD)
%for hadronic observables. The challenge to determine the internal
%structure of nuclei in terms of the distribution of orbital and
%spin angular momentum of their constituents is thus as relevant today
%as it has been at the dawn of nuclear physics.

Knowledge regarding the orbital angular momenta and spin orientations of
the nucleons, bound in the core of an atom, led to a quantitative understanding
of the (hyper)fine structure of the electron shell, \ie, atomic spectra, and
the dynamics of nuclei in external electromagnetic fields. The pioneering
experiments on nuclear magnetic moments were based purely on their electromagnetic
interaction, \eg,  inferring the dependence of resonance frequencies of hydrogen
molecules on an external magnetic field~\cite{PhysRev.56.728}. Such experiments
helped thereby to parametrize nuclear properties in terms of the fundamental constants of
quantum electrodynamics (QED).
The lattice quantum chromodynamics (LQCD)
calculations of the same observables,
\ie, responses of nuclei to external fields, assume analogously
the validity of QCD for nuclei and parameterize them in terms of the constants
of the strong interaction. While both experiment and QCD, in principle,
yield the desired property of every nucleus, clearly not all experiments
nor all LQCD extractions are practical. The predictions which were initially made to fill in
these gaps were based on pairing models for closed-shell nuclei~\cite{PhysRev.46.477}
and required a determination of only the neutron and proton magnetic moments.
Refinements~\cite{PhysRev.69.611} of this model gave insight
to the structural details of few-nucleon wave functions, \eg,
$D$-state admixtures~\cite{PhysRev.70.41}. The fundamental correlation
between nuclear wave functions and electromagnetic responses is part of
the description of nuclei in terms of effective field theories (EFT).
Matching these EFTs to LQCD data is believed to yield a predictive \textit{theory}.

In this article, we apply a candidate for such a theory \eftnopi, as developed in
Refs.~\cite{vanKolck:1998bw,Chen:1999tn,Kaplan:1998we,Kaplan:1998tg,Bedaque:1999ve},
to analyze the structure of two and three-nucleon systems through their
interaction with external electromagnetic probes. The availability of LQCD
calculations at unphysically large quark/pion masses is combined with
experimental data to assess the dependence of charge radii, magnetic moments,
and polarizabilities on nucleon masses, deuteron-triton binding-energy splittings,
and bound states in the two-nucleon singlet channels.
Furthermore, we assess the expected gain in accuracy from dynamical QED,
incorporated into the LQCD extractions of these observables.

% -----------------------------------------------------------------------------
\section{Interaction between nucleons and the electromagnetic field}
\label{Sec:EFTnopi}
%
%%% In this section we shall review the theoretical foundation for
%%% understanding the properties of light nuclei in the presence of
%%% electromagnetic fields. We define the theory by its Lagrangian
%%% and power-counting rules. The ensuing potential is specified
%%% with a set of renormalization conditions which specify the
%%% regularization and input data. Finally, we elaborate on the calculation
%%% of magnetic observables in the Schr\"odinger formalism.

Based on the non-relativistic character of nucleons as constituents of nuclear
bound states, their interaction with external electromagnetic fields and charged
nucleons can be described through a combination of \eftnopi~ with non-relativistic quantum electrodynamics
(NRQED)~\cite{Caswell:1985ui}. The Lagrangian of this effective nuclear theory is
expressed in terms of an iso-spin doublet field $N=\left({p\atop n}\right)$,
which comprises a two-component Pauli spinor for the proton $(p)$ and the
neutron $(n)$, as the most general density conceivable under the constrains
of gauge invariance, locality, hermiticity, parity conservation,
time-reversal symmetry, and Galilean invariance. To leading order (LO) in the
strong interaction and to order
$1/m$ in the Foldy-Wouthuysen-Tani expansion of the Dirac theory, the effective theory, as
relevant for the $A$-nucleon one-photon sector, 
reads \cite{Chen:1999tn}
\ber\label{eq:lagr}
  \mathcal{L} &=& N^\dagger\left\lbrace i\partial_0-e\hat{Q}A_0+
      \frac{1}{2m}\left(\ve{\partial}-ie\hat{Q}\ve{A}\right)^2
     +\hat{g}_N\frac{e}{2m}\ve{\sigma}\cdot\ve{B}\right\rbrace N
   \cr &&
     +\frac{c_T}{m\aleph}\left(N^TP_iN\right)^2
     +\frac{c_S}{m\aleph}\left(N^T\bar{P}_3N\right)^2
     +\frac{d_{3}}{m\aleph^4}\left(N^\dagger\right)^3\left(N\right)^3
   \cr  &&
     +l_1\frac{e}{mm_\pi}\left(N^TP_iN\right)^\dagger\left(N^T\bar{P}_3N\right)B_i
     +l_2\frac{e}{mm_\pi}i\epsilon_{ijk}\left(N^TP_iN\right)^\dagger\left(N^TP_jN\right)B_k\;\;.
\eer
Where, here, and throughout this work, neutrons and protons are assumed to have the same 
(quark mass dependent) mass
$m=m(m_\pi)$. $\ve{A}, \ve{B}$ are the three-dimensional electromagnetic 
vector potential and magnetic fields, 
$\hat{Q}=\frac{1}{2}(1+\tau_3)$ is the charge operator, and
$\hat{g}_N=g_{{p/n}}(1\pm\tau_3)$ the single-particle magnetic moment. 
$P_i$ and $\bar{P}_3$ are projections onto
two-nucleon spin triplet and singlet states, respectively.

Three bare low-energy constants (LECs) $c_{S},c_T,d_{3}$ parameterize the strong
interaction and need to be determined by a matching procedure as well as the four
LECs, $\{g_p,g_n,l_1,l_2\}$, which couple the gauge field to the nucleon(s).
Without its kinetic terms, the radiation field is static. In the Coulomb gauge, the
equation of motion for $A_0$ is time independent and can be integrated to yield
\be
  A_0(\ve{r},t)=e\int\frac{N^\dagger(\ve{r}',t)N(\ve{r}',t)
        +\rho_\text{\scriptsize{ext}}(\ve{r}',t)}{|\ve{r}-\ve{r}'|}d\ve{r}'\;\;,
\ee
where the total charge density in the numerator may contain
dynamical and static ($\rho_\text{\scriptsize{ext}}$) parts.
The former constitutes the Coulomb interaction if substituted in the second
term of the Lagrangian.
Through the static distribution $\rho_\text{\scriptsize ext}$ the single-nucleon
current is coupled to an external charge. Matrix elements of this operator are
usually parameterized by the point-charge radius (see below).
The unnatural scaling of the interaction terms with respect
to a peculiar low-energy scale $\aleph\sim 1/a_s$ ($a_s$ is the scattering length), and a breakdown scale
of the order of the pion mass $m_\pi$ demands a non-perturbative treatment of
the three strong LECs, while the four magnetic couplings are perturbative\footnote{We
assume $e|\ve{B}|\ll mm_\pi\sim 10^{17}~\text{GeV}^2\sim 10^{18}~$G.}.
Of the latter, the two-body parameters $l_1,l_2$ are suppressed by $1/m_\pi$
relative to the one-body terms $g_{n/p}$.
The range of applicability of this theory constrains the momenta of the
interacting nucleons to values below $\sim m_\pi/2$. Within this range, the
Coulomb interaction is non-perturbative for momenta $\lesssim e^2m/4\pi$~\cite{Kong:1999tw}
and requires an additional counter term. For momenta of the order of $e^2m/4\pi$
or larger, \eg, in the helion bound state~\cite{Kirscher:2015zoa,Konig:2015aka}, the interaction is
perturbative. The Lagrangian, subject to these rules, defines \eftnopi~for the
description of light nuclei in the presence of an external magnetic field
and Coulomb-interacting protons.
For practical few-nucleon calculations, we translate the Lagrangian and the
power counting into a nuclear Hamiltonian $\hat{H}_\text{nucl}$
and an interaction Hamiltonian $\hat{H}_{\text{nucl}-B}$
 between the nucleons and the magnetic background field.
\be
  \hat{H}_\text{nucl} = -\sum_{i}^A \frac{\nabla_i^2}{2m} 
                       + \sum_{i<j}^A \hat{V}_{2b}(ij)
                       + \sum_{i<j<k}^A \sum_\text{cyc} \hat{V}_{3b}(ijk) 
\ee
where $\hat{V}_{2b}$, $\hat{V}_{3b}$ are the two and three-body potentials,
\ber
   \hat{V}_{2b}(ij)&=& 
         \left[c^\Lambda_S ~\tfrac{1}{4}(1-\sgmvec_i\cdot\sgmvec_j)
              +c^\Lambda_T ~\tfrac{1}{4}(3+\sgmvec_i\cdot\sgmvec_j)
         \right]\delta_{\Lambda}(\ve{r}_{ij})
         \cr &&    +
         [c^\Lambda_{pp}\delta_{\Lambda}(\ve{r}_{ij})+\frac{e^2}{{r}_{ij}}]
                 ~\tfrac{1}{4}(1+\tauvec_{i,z})(1+\tauvec_{j,z})\;\;,
\eer
and
\be
   \hat{V}_{3b}(ijk)= d_3^\Lambda \delta_{\Lambda}(\ve{r}_{ij},\ve{r}_{ik}) \;\;.
\ee
The regulated delta functions are given by
a gaussian with a smoothing parameter $\Lambda$, 
\ber
  \delta_{\Lambda}(\ve{r}_{ij}) &=& e^{-\frac{\Lambda^2}{4}\ve{r}_{ij}^2} 
  \cr
  \delta_{\Lambda}(\ve{r}_{ij},\ve{r}_{ik}) &=& e^{-\frac{\Lambda^2}{4}\left(\ve{r}_{ij}^2+\ve{r}_{ik}^2\right)}\;\;.
\eer
The interaction between the nucleons and the magnetic field
is exressed through the magnetization density
current
\be
    \hat{H}_{\text{nucl}-B}  
            = (\ve{\mu}^{(1)}+\ve{\mu}^{(2)})\cdot \ve{B}
\ee
where, 
\be
\label{eq:1b}
\ve{\mu}^{(1)}=\sum_{i=1}^A \mu_N\big(\frac{g_p+g_n}{2} \sgmvec_i 
               +\frac{g_p-g_n}{2} \sgmvec_i \tau_{i,z}\big)
\ee
and 
\be
\label{eq:2b}
\ve{\mu}^{(2)}=\sum_{i<j}^A \mu_N   
               \left[
                   l^\Lambda_1(\sgmvec_i-\sgmvec_j)(\tau_{i,z}-\tau_{j,z})
                  +l^\Lambda_2(\sgmvec_i+\sgmvec_j)
               \right]\delta_{\Lambda}(\ve{r}_{ij})\;\;.
\ee
$\mu_N=|e|\hbar/2mc$ is the, $m_\pi$ dependent, natural nuclear magneton (nNM).
The process of eliminating the $\Lambda$ dependence for a set of
observables by absorbing it into the LECs is indicated by the superscripts.
Divergences from the above-mentioned non-local Coulomb repulsion are renormalized
by $c^\Lambda_{pp}$. Like the nucleon mass and the proton charge, the
gyromagnetic factors $g_{p},g_n$ of the nucleons substitute bare LECs.
Projection operators for the two and three-nucleon channels are written
explicitly with standard SU(2) (iso)spin matrices.

To solve the two and three-body Sch\"odinger equation with $\hat{H}_\text{nucl}$
in order to determine bound and scattering states whose properties are used to
calibrate the LECs, and whose $\hat{H}_{\text{nucl}-B}$ matrix elements yield their
leading electromagnetic characteristics, we employ two numerical techniques:
the effective-interaction hyperspherical-harmonic (EIHH)
method~\cite{EIHH1,BARNEA2001565},
and the refined resonating-group (\rgm) method~\cite{Hofmann:1986}.
Details of the numerical implementation of both methods can be found in
Ref.~\cite{Kirscher:2015yda} and references therein.
Besides benchmarking the two numerical techniques, we compare their results with
an analytic two-nucleon calculation in the so-called zero-range approximation
which is identical to \eftnopi~for an infinite regulator $\Lambda$.

Having defined the formal structure and the algorithms used to solve the theory,
we specify observables presumably within its range of applicability in order to, 
first, calibrate the LECs, and second, to exploit its
predictive power.
As in Ref.~\cite{Kirscher:2015yda},
we investigate three different realizations of the
standard model, and thereby the quark-mass dependence of light nuclei.
First, we determine the LECs for the natural (\mpis) case, by matching to experimental data.
The strong interaction parameter $c_T$ is matched to the deuteron binding energy, 
$c_S$ to the neutron-proton-singlet scattering length,
and $d_3$ to the triton binding energy. The magnetic parameters are matched
to the magnetic moments of the triton $(l_1)$ and the deuteron ($l_2$).
Second, we match to lattice QCD predictions for SU(3)-degenerate quarks with a mass
corresponding to \mpil. In this case there is a bound 
two-nucleon-singlet state. Therefore, $c_S$ is adapted to reproduce its binding energy,
and the magnetic parameter $l_1$ is adapted to the transition matrix
element between the singlet and triplet two-nucleon states
$t_{01}=\bra S=1 ~| \hat{\mu}_z|~ S=0 ~\ket $. 
All other LECs are fitted to the same
observables as at physical pion mass.
For the intermediate pion mass \mpim ~there are
available two-nucleon LQCD
binding energies, which we utilize to
constrain $c_S,~c_T$. For the magnetic couplings $l_{1},l_2$,
we interpolate values between those at physical and 806~MeV pion mass. This
interpolation is analogous to that in Ref.~\cite{Beane:2015yha} where it was used to
predict the
radiative-capture cross section $np\to d\gamma$ at physical $m_\pi$ from
data at 450~MeV and 806~MeV pion mass. 
%%% While an empirical $m_\pi$ insensitivity
%%% of the magnetic moments (in units of natural nuclear magnetons) motivates
%%% this interpolation, we calculate the dependence of the triton
%%% magnetic moment on its binding energy. As the $m_\pi$ insensitivity reflects the
%%% quality of the shell model of the triton, the binding-energy dependence of the moment
%%% should be equally weak.
The available constraining observables are listed in Table~\ref{tab:input} along with their numerical
values.
% --
\renewcommand{\arraystretch}{0.75}
\begin{table}[h]
\begin{center}
\caption{\label{tab:input}\small Experimental and LQCD data for
Binding energies ([MeV]), magnetic moments ([nNM]), the two-body transition matrix element
$t_{01}$ ([nNM]),
and scattering lengths ([fm]). 
}
\begin{tabular}
{c @{\hspace{5mm}} c @{\hspace{5mm}} c @{\hspace{5mm}} c @{\hspace{5mm}} }
\hline\hline
   Observable  & Nature~\cite{RevModPhys.84.1527} \mpis & LQCD \mpim\cite{Beane:2015yha,Orginos:2015aya,Parreno:2016fwu} & LQCD \mpil\cite{Beane:2012vq,Chang:2015qxa} \\
\hline
$m$      & $938.9$ & $1226(12)$ & $1634(18)$ \\
\hline
$\mu_n$    & $-1.913$ & $-1.908(38)$ & $-1.981(19)$ \\
$\mu_p$    &  $2.793$ &  $2.895(56)$ &  $3.119(74)$ \\
\hline
$B_{np}$ & $-$     & $12.5(50)$  & $15.9(40)$\\
$a_{np}^\text{\tiny singlet}$ & $-23.75$ & - & - \\
$a_{pp}$  & $-7.806$ & - & - \\
$\bd$    & $2.225$   & $14.4(32)$   & $19.5(48)$ \\
$\mud$   & $0.857$   & $-$          & $1.22(10)$ \\
$t_{01}$  & $-$ & $-$ & $5.48(20)$ \\
\hline
$\bt$    & $8.482$   & $-$ & $53.9(107)$ \\
$\mut$   & $2.979$   & $-$ & $3.56(19)$ \\
$\bh$    & $7.718$   & $-$ & $-$ \\
$\muh$   & $-2.127$  & $-$ & $-$ \\
\hline\hline
\end{tabular}
\end{center}
\end{table}
\renewcommand{\arraystretch}{1.00}
% --

A comment about the Coulomb interaction between protons is in order. 
While the proton-proton scattering length and the $^3$He binding energy are
known experimentally, LQCD calculations which consider some version of QED
for the electromagnetic interaction of the quarks are, as of now, unattainable.
In order to estimate the effect of dynamical U(1) gauge fields, we proceed as
follows. We assume that QCD corrections to the QED fine-structure constant
$\alpha$ are insignificant for the accuracy of this work.
What justifies the perturbative treatment of the Coulomb
force for physical $^3$He holds also for the bound two and three-nucleon states
containing two protons, \ie, the $pp$ singlet, and $^3$He with heavier pions.
These systems should even be more amenable to a perturbative expansion because
of the larger binding momenta associated with their binding energies
(Table~\ref{tab:input}).
An ansatz for the effective interaction resultant from quark QED as a Coulomb exchange,
whose iterations should be strongly suppressed in bound states, and a counter term to
renormalize low-energy amplitudes seems appropriate. 
We expect this ``model'' to shift the di-proton binding energy by the
amount an iterated Coulomb interaction with $\alpha=\alpha_\text{physical}$ determines,
plus a correction from $c_{pp}$ to eliminate cutoff dependence. We fixed $c^\Lambda_{pp}$
by enforcing the split \mbox{$B(np)-B(pp)=0.5~$MeV}.
As this differs from a splitting induced by Coulomb by $\ll 1~$MeV
over the considered cutoff range (see discussion of Fig.~\ref{fig:cstpp}), we
cannot discriminate the ensuing $c_{pp}$ from other values which set the splitting
at values which differ by $\sim1~$MeV. All choices for the splitting correspond to
different effective QED models of which we assess two, $c_{pp}$ to yield
the $0.5$~MeV splitting and $c_{pp}=0$ to yield an insignificantly $\Lambda$-dependent splitting
$\lesssim 1~$MeV.
% -----------------------------------------------------------------------------
%\input{joki_results_v2}
\section{Results}
\label{Sec:RES}
The EFT defined above is utilized to pre/postdict electromagnetic characteristics
of the proton-proton, the singlet-neutron-proton, the deuteron, triton, and helium
systems in the form of point-charge radii, magnetic moments, and magnetic
polarizabilities. Numerical results are compiled in Table~\ref{tab:pred} as
obtained for the three pion masses where enough data is available to renormalize
the EFT. The uncertainties are to be viewed as lower bounds as they are
inferred solely from the $\Lambda$ sensitivity. For the consistency analysis
discussed in subsection~\ref{sec:resA}, we also considered the uncertainty in the
input data but used the central LEC values for subsequent calculations.
% --
\renewcommand{\arraystretch}{0.75}
\begin{table}[h]
\begin{center}
\caption{\label{tab:pred}\small \eftnopi~results ($\Lambda\to\infty$ extrapolations)
for point-proton charge radii
($r_\text{ch}\equiv\bra r_p^2~\ket^{1/2}~$[fm]),
magnetic moments ([nNM]), and 
polarizabilities ([fm$^3$]).
Preexisting experimental~\cite{RevModPhys.84.1527} or LQCD values~\cite{Chang:2015qxa} are 
written below EFT postdictions. Single
entries represent true EFT predictions. 
}
\begin{tabular}
{l @{\hspace{5mm}} r @{\hspace{5mm}} c @{\hspace{5mm}} c @{\hspace{5mm}} c @{\hspace{5mm}} }
\hline\hline
   \multicolumn{2}{c}{}  & \mpis & \mpim & \mpil \\
\hline
%proton-proton  
%           & $r_\text{ch}$ & $-$             & $0.825(290)$   & $0.647(260)$\\
NN-singlet 
           & $r_\text{ch}$ & $-$             & $0.588(260)$   & $0.458(240)$\\ \hline
deuteron   & $r_\text{ch}$ & $1.55(24)$      & $0.550(250)$   & $0.416(250)$\\
           &              & {Exp. } $1.97$  &                & \\ \cline{2-5}
           & $\beta_M$    & $0.0733(1)$     &  &  $2.92(1)~10^{-2}$      \\ 
%%%           &              & {sum rule~\cite{Gorchtein:2015eoa}} $0.072$ &  &  {LQCD}  $4.4(17)~10^{-4}$     \\ 
           &              & {sum rule~\cite{Gorchtein:2015eoa}} $0.072$ &  &   \\ 
           &              & {AV18~\cite{FriarPayne97-2}} $0.0774$ &  &   \\ 
           &              & {EFT~\cite{CHEN1998221}}            $0.096$ &  &   \\ \hline
%-----------------------------------------------------------------------------
triton     & $r_\text{ch}$ & $1.16(23)$    & $0.767(310)$   & $0.460(280)$    \\
           &              & { Exp.~\cite{Angeli201369}} $1.55$          & & \\
           &              & { LO-EFT~\cite{Vanasse:2015fph}} $1.13(34)$ & & \\ \cline{2-5}
           & $\mu$        & $2.9710$        & $3.08(6)$      & $3.41(3)$       \\
           &              & { Exp.} $2.979$   &  & { LQCD}  $3.56(18)$   \\ \cline{2-5}
           & $\beta_M$    & $8.2(1)~10^{-4}$ & $-$ & ${3.9(4)~10^{-5}}$  \\ 
           &              &                 &     & { LQCD } $2.6(18)~10^{-4}$  \\ \hline
%-----------------------------------------------------------------------------
helion     & $r_\text{ch}$ & $1.30(28)$      & $0.793(300)$   & $0.472(290)$ \\
           &              & {Exp. } $1.78$  &                &              \\ \cline{2-5}
%          & $B$          & ${7.64(5)\atop\text{\tiny $7.718$}}$     & $28.7(5)$      & $52.3(7)$\\
%          & $B$          & ${7.64(5)\atop\text{\tiny $7.718$}}$     & $28.7(5)$      & $52.3(7)$\\ \cline{2-5}
           & $\mu$        & $-2.13(1)$      & $-2.15(5)$     & $-2.17(6)$\\
           &              & {Exp.} $-2.127$ &     & {LQCD } $-2.29(12)$\\ \cline{2-5}
           & $\beta_M$    & $9.7(1)~10^{-4}$ &  & $3.9(4)~10^{-5}$  \\ 
           &              &                 &  & {LQCD } $5.4(21)~10^{-4}$  \\ 
\hline\hline
\end{tabular}
\end{center}
\end{table}
\renewcommand{\arraystretch}{1.0}
% --
%
\subsection{Low-energy constants and data consistency}
\label{sec:resA}
\begin{figure}[h]
  \includegraphics[width=1.0 \textwidth]{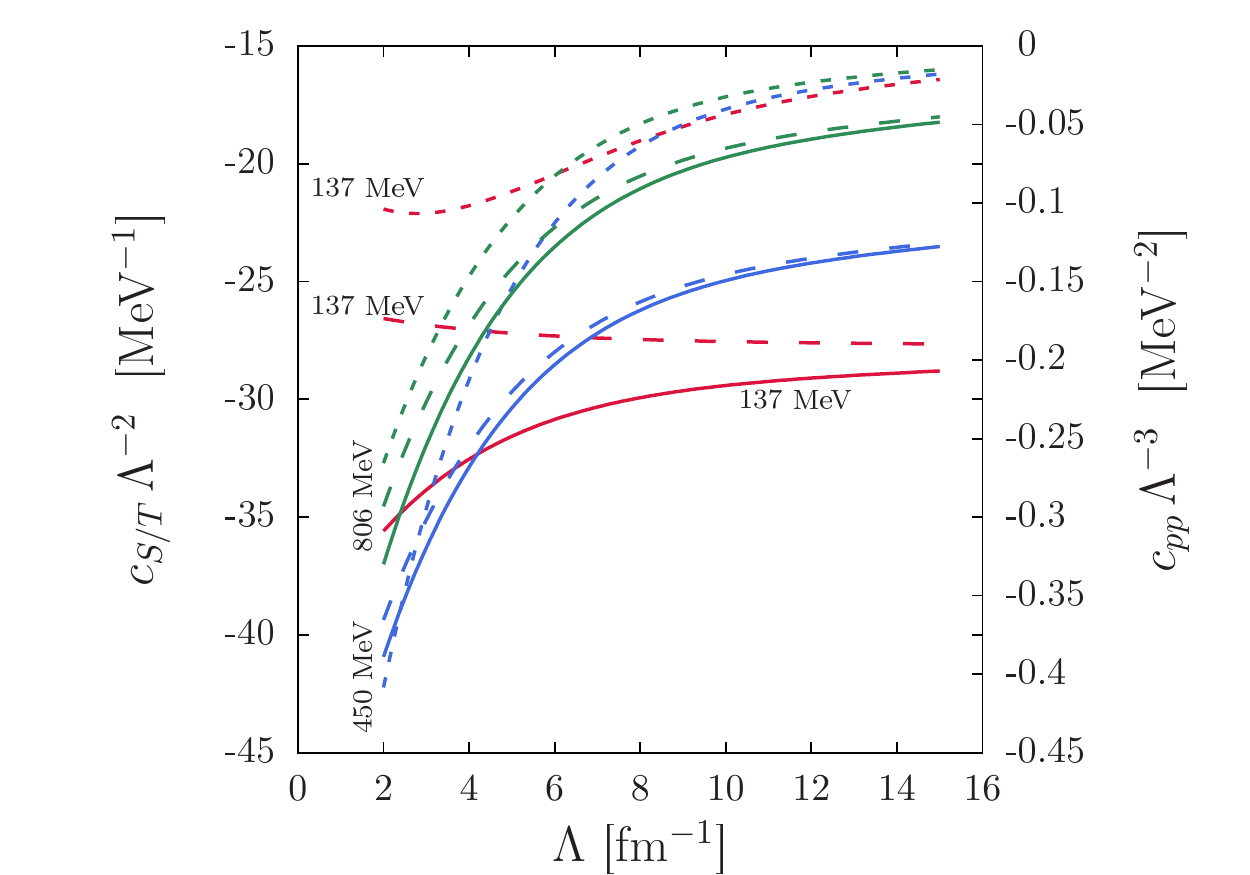}
  \caption{\label{fig:cstpp}(Color online) Cutoff dependence of the LECs $c_S$
  (solid line, left y-axis), $c_T$ (dashed line, left y-axis), and $c_{pp}$ (short dashed, right y-axis)
  for three pion masses, \mpis~(red), \mpim~(blue), and \mpil~(green).}
\end{figure}

The renormalization of the EFT demands regulator independence
of a set of observables. With this set taken as specified in the previous section,
we obtain a cutoff dependence of the LECs as shown in
Fig.~\ref{fig:cstpp} for $c_S,c_T,c_{pp}$, and Fig.~\ref{fig:l1l2} for $l_1,l_2$.
The numerical values of these LECs are presented in Appendix \ref{sec:lecstbl}.
For a thorough discussion of the behavior of $c_{S},c_T$, namely, the dominating
$\Lambda^2$ dependence and the small Wigner-SU(4)-symmetry breaking component
(overlapping solid and dashed lines in Fig.~\ref{fig:cstpp}
for $\Lambda\to\infty$), we
refer to Ref.~\cite{Kirscher:2015yda}.
A different dependence
of the small correction term $c_{pp}$ in the proton-proton channel is found here:
an asymptotic behavior (short dashed lines, right y-axis
in Fig.~\ref{fig:cstpp}) 
for all three pion masses of $\lim\limits_{\Lambda\to\infty}c_{pp}\propto\Lambda^3$.
This unmasks the difference of the divergence structure of the Coulomb exchange as
found in Ref.~\cite{Kong:1999tw} relative to that of a two-nucleon loop. The latter is
absorbed into $c_{S},c_T$, while $c_{pp}$ is needed if the bubble is cut by a
static Coulomb exchange.

A comment about previous calculations which demand $c_{pp}$ is in order. Here, we find
$c_{pp}$ to adjust $c_S$ by less than $0.1\%$
(compare scales in Fig.~\ref{fig:cstpp})
over the considered cutoff range from 2~fm$^{-1}$ to 15~fm$^{-1}$. Despite
the enhanced effect on observables, setting $c_{pp}=0$, as in Ref.~\cite{Kirscher:2009aj},
does not indicate a severe cutoff dependence, \eg, in predictions for the
$^3$He binding energy or the proton-proton scattering length. This fallacy
is a consequence of the specific regularization chosen here, and was avoided in, \eg,
Ref.~\cite{Vanasse:2014kxa} with a different scheme, and in Ref.~\cite{Kirscher:2015zoa} with
the same formalism as employed in this work.
Within our scheme, we find the divergence only by splitting the LEC in the
$pp$ channel as shown. 
%%% The $\Lambda$ dependence of the ensuing $c_{pp}$ is
%%% then found almost identical for all three pion masses, despite the fact,
%%% that at \mpis~it is calibrated to the scattering length and
%%% for the other two scenarios to the $pp$ binding energy.
%--
\begin{figure}[h]
  \includegraphics[width=1.0 \textwidth]{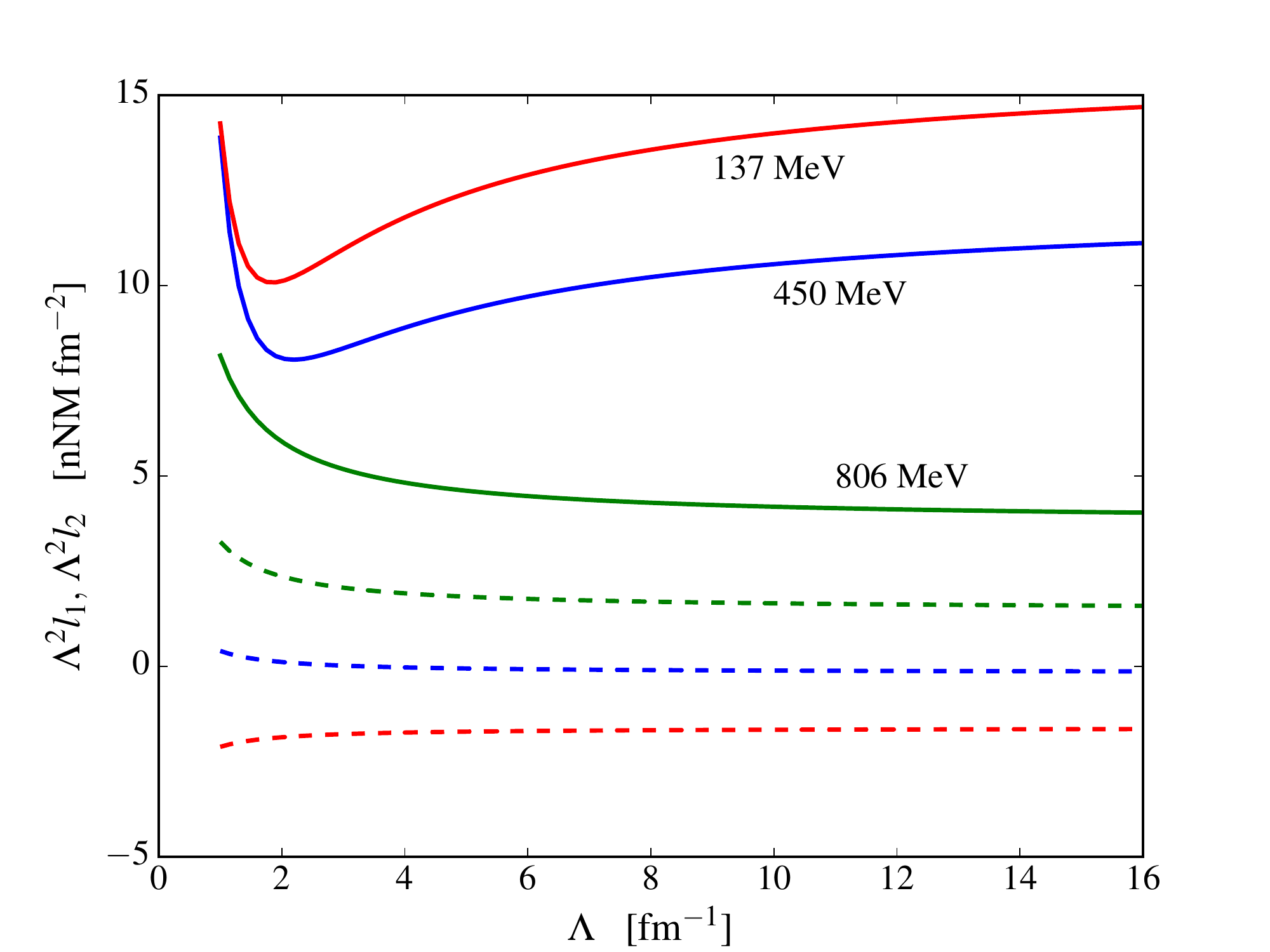}
  \caption{\label{fig:l1l2} (Color online) Cutoff dependence of the LECs $l_1$
  (solid line), and $l_2$ (dashed line) 
  for three pion masses, \mpis~(red), \mpim~(blue), and \mpil~(green).
  The values for \mpis, and \mpil~ are fitted to experimental and LQCD date respectively.
  The \mpim~ values are results of an interpolation.}
\end{figure}
%--

For the coupling of the photon to the two-nucleon vertex, \ie $l_{1},l_2$,
we observe an asymptotic behavior of
$\lim\limits_{\Lambda\to\infty}l_i\propto\Lambda^{-2}$.
This dependence can be derived analytically by understanding the limit
$\Lambda\to\infty$ as the well-known zero-range approximation (see Appendix~\ref{sec.app1}). 
The relatively slow convergence and the different
behavior at small $\Lambda$ of $l_1$ at \mpil~
(green solid, Fig.~\ref{fig:l1l2})
is consistent with previous findings~\cite{Kirscher:2015yda}, which already showed the
necessity of larger cutoffs for this large pion mass due to the associated
large binding momenta. Another peculiarity at the largest pion mass is the sign
difference of $l_2$ compared to the physical point. This is understood
from the comparison of the deuteron's magnetic moment to those of its
constituents. At leading order, $\mud=\mu_p+\mu_n$, which is larger than
the experimental value but smaller than the lattice measurement at \mpil.
The next-to-leading-order (NLO) $l_2$ term thus either reduces or enlarges $\mud$.
%-----------------------------------------------------------------------------
\begin{figure}[h]
\includegraphics[width=1.0 \textwidth]{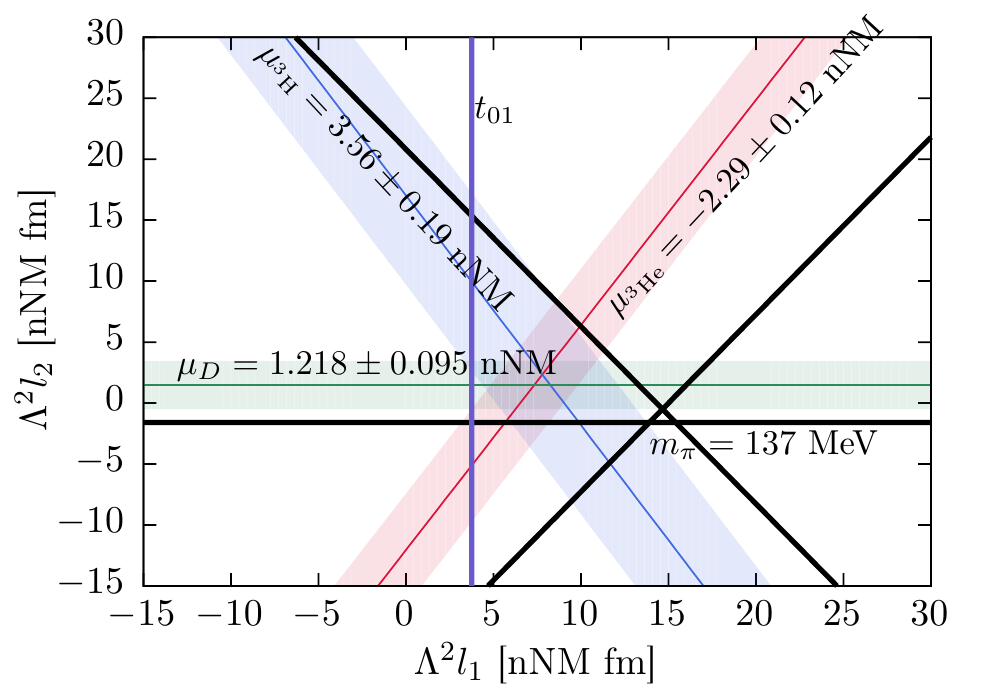}

\caption{\label{fig:lec-cons} \small (Color online) Interdependence of two-body LECs
consistent with the magnetic moments of the deuteron (green, horizontal),
the triton (blue, negative slope), $^3$He (red, positive slope), and the
magnetic-field contribution to the di-nucleon energy splitting
$\delta E_{^3S_1,^1S_0}$ (purple, vertical) at $m_\pi=806~$MeV.
Bandwidth resembles the total lattice uncertainty in the respective
observable.
The black lines marks the LEC values which yield the experimental
deuteron, triton, and 3-helium magnetic moments at the physical pion mass.
}
\end{figure}
%-----------------------------------------------------------------------------
To attest to the consistency of the theory with the measured and calculated data,
we compare possible matching conditions on $l_1$ and $l_2$ in Fig.~\ref{fig:lec-cons}.
Each band shown in the figure defines the area of allowed $l_1,l_2$ pairs, which
are consistent with one measurement/calculation of a magnetic moment. As $\mud$
is insensitive to the $l_1$ term, it only constrained $l_2$. This constrain is
shown by an horizontal
band, with a width representing the total uncertainty where we considered 
statistical and systematic errors in quadrature. 
At larger pion masses, an electromagnetically induced transition between the singlet
and triplet bound states is allowed. The respective matrix element has been calculated
with LQCD, and we can constrain the EFT with this additional input,
$t_{01}$. 
At physical $m_\pi$, this transition represents a breakup or fusion of a deuteron or a scattering
neutron-proton singlet, respectively. 
%%% As the associated cross sections are affected by
%%% other transition, too. 
This constraint is not used here at physical $m_\pi$.
The lattice predictions for
$\mut$ ($\muh$) constrain the LECs to a negatively (positively) sloped 
band. The slope $dl_2/dl_1$
has the same magnitude but opposite sign, dependent upon whether $\mut$ 
or $\muh$ is used as a constraint. This follows from the structure
of the $l_1$ operator (Eq.~\ref{eq:2b}) whose isospin matrix element
flips sign, while spin and coordinate-space matrix elements are identical
at 806~MeV and almost equal at physical $m_\pi$.

Consistency between data and theory is attested in Fig.~\ref{fig:lec-cons} by an
overlap region of all four bands. The $l_2(l_1)$ dependencies shown in the
figure are for extrapolations $\Lambda\to\infty$ from the interval
$4-15~\text{fm}^{-1}$ in which the necessary matrix elements were obtained.
The EFT uncertainty is not explicit in the graph, but it is
responsible for the three physical lines not intersecting in a point.
In the \mpil~ case, we see a similar situation considering constraints due to magnetic moments.
On the other hand, the transition matrix element $t_{01}$ seems to be inconsistent with the other 
observables, although still 
acceptable since it is within the current LQCD error bars.
 
%\com{For Johannes: maybe some comments about non findings, moments of not the deep state
% but something else, Evgeny's LET analysis.}

%-----------------------------------------------------------------------------
%\input{./fig_ebmu}
\begin{figure}[h]
\includegraphics[width=1.0 \textwidth]{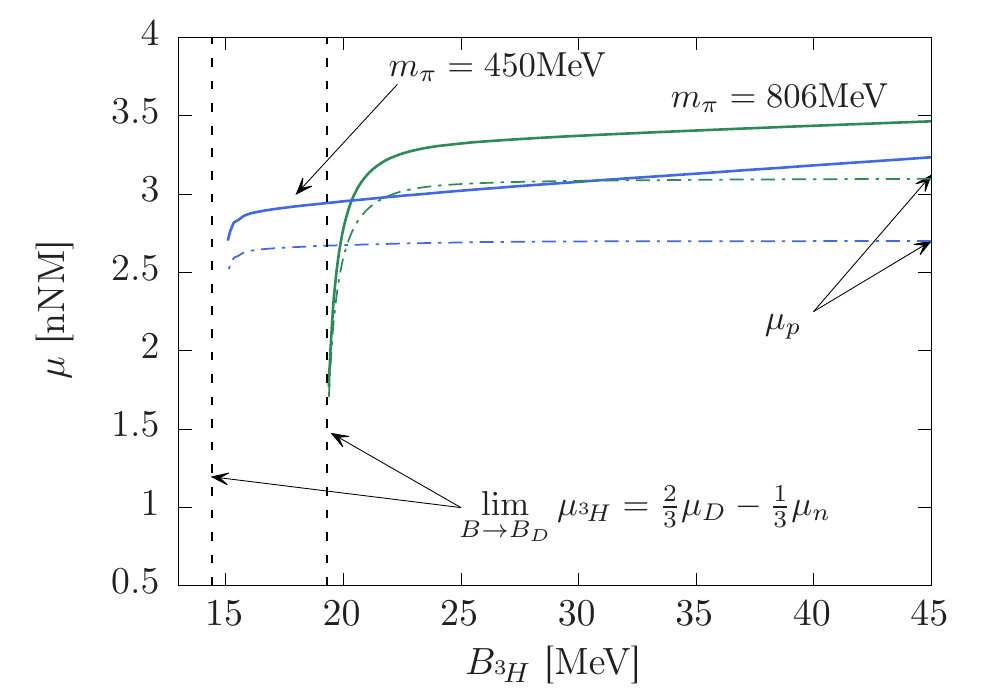}
\caption{\label{fig:ebmu}\small (Color online) The magnetic moment of the triton as a function of its binding energy for \mpil~(green, $\Lambda=15~$fm$^{-1}$)
and \mpim~(blue, $\Lambda=8~$fm$^{-1}$).
Vertical dashed lines mark the deuteron-neutron thresholds at
$B_D=19.5~$MeV and $B_D=14.4~$MeV, respectively. LO results with
one-body-current coupling (dash-dotted lines) are compared with NLO
values (solid lines) which consider also the two-body-current coupling $l_{1},l_2$.
Asymptotic limits are indicated with arrows,
for $\bt\to\bd$: $\mut\to1.196~$nNM (450~MeV), $\mut\to1.472~$nNM (806~MeV);
and $\bt\to\infty$: $\mut\to2.70~$nNM (450~MeV), $\mut\to3.119~$nNM (806~MeV).
}
\end{figure}
%%-----------------------------------------------------------------------------
\subsection{Three nucleons at \mpim}
\label{sec:res450}
We begin the discussion of observables at the pion mass
where not enough data has been calculated to calibrate all LECs,
and we rely on interpolated values for $l_{1},l_2$, as described above.
For predictions in the three-nucleon sector one three-body observable is
required to renormalize the EFT. No such datum has been calculated
at \mpim. The magnetic moment of the triton, for example, can thus only be
given as a function of its binding energy. This dependence is shown
in Fig.~\ref{fig:ebmu} for the two unphysical pion masses. Results at LO 
and NLO in the coupling of the magnetic field are shown. For $\bt$ slightly
larger than the threshold energy $\bd$, the LO dependencies converge to a constant,
while at NLO, $\mut$ rises linearly with $\bt$. In the limit of $\bt\to\bd$, \ie,
for barely bound, very shallow states, all curves approach the na\"ive limit
$\mut\sim2/3\mud-1/3\mu_n$ of a free deuteron-neutron system with appropriate 
spin orientation. In the other limit, $\bt\to\infty$, LO results are identical to
the shell-model/Schmidt~\cite{Schmidt1937} values and thus provide a \textit{deep}
consistency check for the numerical method to produce the compact triton.
The deviation $\delta\mut$ from the Schmidt limit due to the photon coupling to the
two-nucleon contact is about $15\%$ and vanishes only at threshold. Above some
critical binding energy, about 2-4 MeV above threshold, $\delta\mut$ changes linearly with
$\bt$.

Assuming that $3/2\bd(450)<\bt(450)<\bt(806)$, the correlation in Fig.~\ref{fig:ebmu}
yields the constraint:
\be
\mut=3\pm0.3~\text{nNM}\;\;\text{at}\;\; m_\pi=450~\text{MeV}\;\;.
\ee
A linear interpolation
between $\bt$'s at physical and 806~MeV $m_\pi$ suggests a central value of $\bt=29.7~$MeV.

%=============================================================================
\subsection{Charge radii}
%=============================================================================
We shall employ the theory now to analyze
the spatial distribution of nucleons within a nucleus
at all three pion masses. Canonically,
this is encoded in the radial moments of a nucleus. These moments 
are expansion coefficients of form factors. We consider the coupling of a nucleus to
an external electric charge distribution which is parameterized with a charge form factor 
\be
\label{eq:ff_exp}
F_C(q^2)=1-{\bra r_p^2 ~\ket \over 6}q^2 +\ldots\;\;.
\ee
It is implicit in this expansion that
the Lagrangian Eq.~\ref{eq:lagr} does not contain a coupling of the external
charge to a four-nucleon vertex. Thus, it suffices to consider the one-body, scalar
coupling via $\rho_\text{\scriptsize{ext}}$ (Eq.~\ref{eq:lagr}), analog to the leading
contribution to the magnetic moment (see below).
Two-body-current contributions to the charge radius appear at $O(Q^3)$ as
described in Ref.~\cite{Valderrama:2014vra}, and thus 
the point-charge radius calculation
for an $A$-nucleon bound state with $Z$ protons amounts to:
\be\label{eq:charge_rad}
  \bra r_p^2 ~\ket={1 \over Z} \bra A |\sum_{i=1}^A {1 \over 2}(1+\tau_{z,i}){\bf r}_i^2 | A ~\ket\;\;.
\ee
We obtain the bound-state wave function as a solution of the Schr\"odinger equation in coordinate space
with the above defined interaction. Nucleons are assumed to be point-like in this approach, and hence
the comparison with experiment becomes more favorable if the datum, the charge radius
$\bra r_\text{c}^2 ~\ket$, is
corrected by a finite proton and neutron size\footnote{$R_p\approx0.841~$fm, and $R_n\approx-0.116~$fm, respectively.}, $\bra r_c^2 ~\ket=\bra r_p^2 ~\ket
+R_p^2+N/(A-N)R_n^2$. 
%%% However, these terms, and others which are sometimes
%%% included to account for recoil or meson-exchange effects when model calculations are
%%% compared with data, must emerge order by order in the EFT expansion. In an EFT approach,
%%% these terms are part of the theoretical uncertainty, and removing them by hand is unjustified.
%%% Hence, we list experimental values for the charge, instead of the point-charge
%%% radius in Table~\ref{tab:pred}.
%-----------------------------------------------------------------------------
\begin{figure}[h]
\includegraphics[width=1.0 \textwidth]{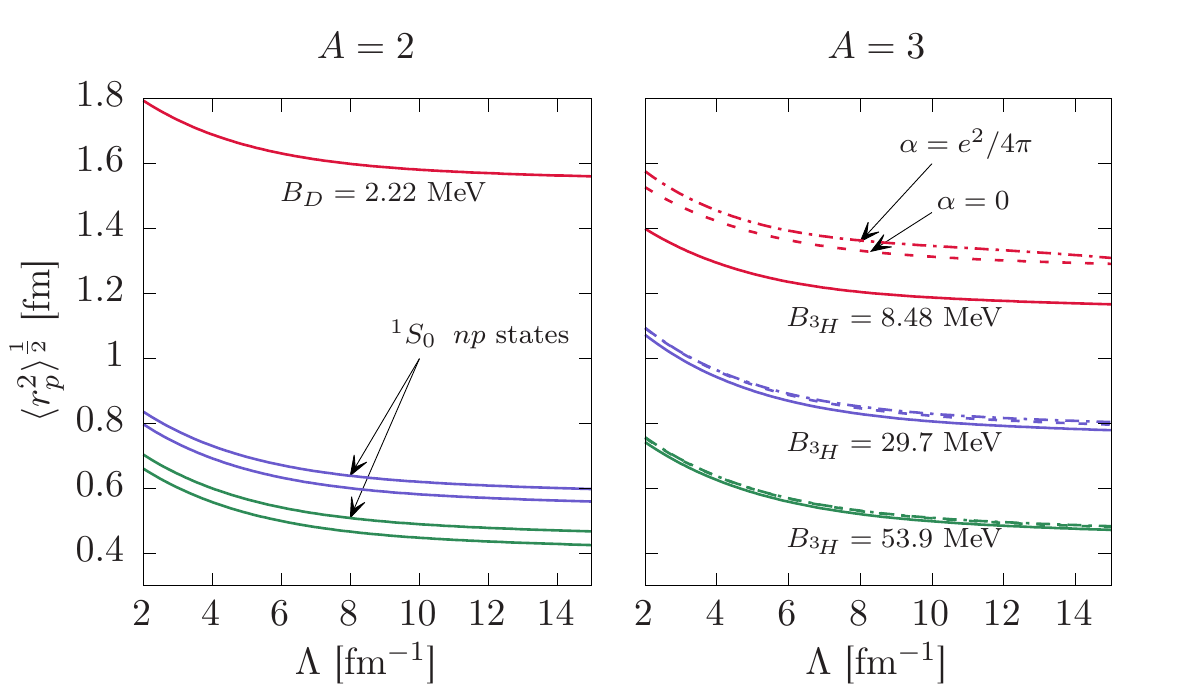}
\caption{\label{fig:radii}\small (Color online) Regulator dependence of the point-charge
radii for the two (left panel) and three-nucleon (right panel) bound states
at the physical (red), 450~MeV (blue), and 806~MeV (green) pion mass.
Solid lines refer to one-proton systems, \ie, the deuteron, $np$, and triton.
Results for two-proton systems are shown with (dash-dotted) and without (dashed)
electrostatic repulsion between the protons.
For $A=2$, the lines indicated with an arrow
correspond to singlet $np$ state, while the lower solid
lines mark the triplet deuteron.}
\end{figure}
%%-----------------------------------------------------------------------------
\paragraph*{The $A=2$ case -} 
The dependence on the Gaussian regulator for all two-nucleon
bound states at the physical and two unphysical pion masses is
given in the left panel of Fig.~\ref{fig:radii}.
We find approximately the same $\Lambda$-convergence rate for the radii of the deuteron,
and the singlet $np$.
In turn, the difference between the respective values is $\Lambda$ independent which
reflects the variation in the binding energies,
that are cutoff independent by construction.

The $np$ singlet states at larger pion masses are not as deeply bound as the triplet
states. A binding-energy difference of $\delta B_{450}\sim 1.9~$MeV
and $\delta B_{806}\sim 3.6~$MeV, respectively, results in charge radii which are
different by an amount smaller than the EFT uncertainty\footnote{A lower bound of which
is given by the difference of the radii obtained at smallest and largest $\Lambda$,
\ie, about $0.3$~fm (see Table~\ref{tab:pred}).}. 
With no electromagnetic repulsion between the protons, the charge radii of the proton-proton
and neutron-proton singlets are identical. 
Even the effect of a Coulomb-induced splitting \mbox{$B(np)-B(pp)=0.5~$MeV}
(see discussion of $c_{pp}$ calibration) is found insignificant,
\ie, $\bra r_\text{c}^2 ~\ket$ of the now shallower di-proton
is still almost identical to that of the $\alpha=0$ scenario. 
Based on this observation, one would not expect LQCD predictions at 450~MeV and
806~MeV $m_\pi$ of this observable to be affected strongly by dynamical QED.

%This insensitivity suggest that a change in size is not driven 
%primarily by the binding energy of the system.
%On the one hand, \mbox{$B(np)-B(pp)=0.5~$MeV}
%is less than 20\% of the binding-energy splitting between the singlet and triplet $np$ states.
%The change in the radius, on the other hand, is large (dash-dotted compared to solid lines).
%Second, the effect of the electric interaction between the protons is apparently
%insignificant as dashed and dash-dotted lines are found on top of each other
%(left panel Fig.~\ref{fig:radii}).

%-----------------------------------------------------------------------------
\paragraph*{The $A=3$ case -}
The $\Lambda$ dependencies of the point-charge radii of the triton 
and $^3$He (Fig.~\ref{fig:radii})
suggest again approximately equal theoretical EFT uncertainties for all pion masses,
as inferred from the shape similarity of the respective curves.
Again, the main motivation for this analysis is to assess the sensitivity of the
observable with respect to electromagnetic interactions between the nucleons. At
\mpis, the additional proton in $^3$He results in a significantly
larger system, even if no Coulomb interaction is included. Note the
difference to the two-nucleon case, where energetically degenerate $pp$ and $np$ singlets
do also have the same charge radius. For three nucleons, an identical binding energy
for the triton and $^3$He: $8.48~$MeV, does not produce the same charge radii. 
The effect of the Coulomb repulsion and the $c_{pp}$ counter term, which is
adjusted to the $pp$ scattering length, is relatively small, yet seizable (dashed and
dash-dotted lines). At \mpim, the respective differences in the radius between the triton
and the charged and uncharged $^3$He are tiny. Finally, at \mpil, all three systems
yield almost identical point-charge radii.

The results do not identify the binding energy as the main factor inducing the
differences in this observable. This is apparent at physical $m_\pi$, where the uncharged $^3$He 
has the same binding energy as the triton. The latter is $\Lambda$ independent
by construction while the binding energy of the charged $^3$He nucleus
is subject to a theoretical uncertainty within the considered $\Lambda$ range
because it is the $pp$ scattering length (\mpis) or the $pp$ binding energies
(unphysical $m_\pi$'s) which are used to renormalize $c_{pp}$. This residual
$\Lambda$ dependence of $\bh$ is not reflected in the results as we find the shape of the
corresponding dash-dotted curves in the right panel of Fig.~\ref{fig:radii}
indistinguishable from those which represent systems with fixed binding energy.

In our analysis, we therefore idetify the breaking of the Wigner $SU(4)$ symmetry,
as the main source of this differnce in the point-charge radii of $^3$H and $^3$He.
For an $SU(4)$ symmetric triton or helion we would
expect the neutron point-charge radius to be identical to the proton point-charge radius, 
and to the matter
radius. The breaking of this symmetry enlarges the radius of the majority species, since the $^1S_0$ channel is
less attractive then the $^3S_1$ channel. 
At higher pion masses \cite{Kirscher:2015yda}
the $SU(4)$ symmetry is restored, and as a consequence we see the point proton charge radius difference shrinking
with increasing pion mass. 

The conclusion is the same as in the two-nucleon sector: the QED
uncertainty in LQCD predictions of this observable at the large pion masses
is expected to be negligible.
\paragraph*{Comparing the $A=2~\&~3$ cases -}
A comparison of radii in two and three-nucleon systems supports the refutations of a
correlation between system size, as measured by the point-charge radius,
and binding energy. At \mpis, this correlation would still yield the correct
hierarchy with the triton as the most deeply bound, and thus smallest, system, followed
by $^3$He, which is less deeply bound and larger, up to the largest and shallowest
deuteron. In contrast, we find all three-nucleon systems larger in size at the
unphysical pion masses relative to the $np$ bound states, despite the fact that the
latter are much less deeply bound. At \mpim, two and three-nucleon systems have
the same charge radius given the uncertainties. The counter-intuitive ordering of
two and three-nucleon radii is a first indication of the peculiarity of the
NN system at \mpil. 
%%% Below, we will encounter the polarizability as another.
In conclusion of this section, we note that the orderings are unaffected
by the regularized Coulomb interaction and consequently should be characteristics
of the strong interaction. 
%=============================================================================
\subsection{Magnetic moments}
%=============================================================================
\renewcommand{\arraystretch}{0.75}
\begin{table}[h]
\begin{center}
\caption{\label{tab:magmom} The evolution of the magnetic moments (in [nNM]) of 
the $A=2,3$ nuclei in EFT($\nopi$) for \mpis, and \mpil.
}
\begin{tabular}
{l @{\hspace{5mm}} c @{\hspace{5mm}} c @{\hspace{5mm}} c @{\hspace{5mm}} 
 c @{\hspace{5mm}} c @{\hspace{5mm}} c @{\hspace{5mm}} }
\hline\hline
   & \multicolumn{3}{c}{\mpis} & \multicolumn{3}{c}{\mpil} \\
   & deutron & triton & helion    & deutron & triton & helion \\
\hline
  shell model & 0.879   & 2.793  & -1.913 & 1.138     & 3.119 & -1.981  \\
  LO          & 0.879   & 2.746  & -1.862 & 1.138     & 3.118 & -1.979  \\
  NLO         & 0.857   & 2.979  & -2.130 & 1.220     & 3.405 & -2.170  \\ \hline
  EXP/LQCD    & 0.857   & 2.979  & -2.127 & 1.220(95) & 3.56(19) & -2.29(12) \\
\hline\hline
\end{tabular}
\end{center}
\end{table}
\renewcommand{\arraystretch}{1.00}

In Table \ref{tab:magmom} we present the evolution of the nuclear magnetic moments in EFT($\nopi$).
The values of shell-model approximation yield the magnetic moment
as the sum of the single particle contributions with appropriate spin orientations.
This simple approximation works well within 15\%~for \mpis, and \mpil, for all considered nuclei. 
We then consider the coupling of the LO \eftnopi~ magnetic one-body currents to a bound nucleus, 
as first refinement of the shell model.
%At the next step we include a more detailed description of the nuclear wave 
%function within EFT($\nopi$), utilizing only LO 1-body currents.
As expected, the deuteron magnetic moment is unaffected. However, the
agreement between theory and data gets worse for the $A=3$ nuclei, particularly at the 
physical pion mass.
To understand this result, we should return to the discussion in \ref{sec:res450}~
and consider the competing pictures of a compact $A=3$ nucleus versus a 
shallow cluster state composed of a neutron or proton orbiting around a 
deuteron. For a compact nucleus, the single-particle picture:
$\mut=\mu_p$, and $\muh=\mu_n$,  dominates. For a clustered state,
we expect that $\mut\longrightarrow (2/3\mud-1/3\mu_n)$ as $\bt\longrightarrow\bd$, and therefore
to obtain a smaller magnetic moment (this argument applies equally to $^3$He).
This explanation is consistent with the difference in binding energies
between the rather shallow trimers at the physical pion mass, 
and the deeply bound \mpil\; trimers.

The two-body magnetization current that appears at NLO, reconciles the theory with 
the available data.
For the physical case we see an agreement at the 2 permil level. This might not be that
impressive as $l_1$ was fitted to reproduce the $^3$H magnetic moment. In contrast 
the $A=3$ results for the \mpil ~case are prediction of our theory, and it can be seen 
that they agree with the LQCD data within error bars.

The discrepancy between the nuclear magnetic moments and theoretical predictions
relaying on the one-body magnetization current, only, have a history in nuclear physics.
It was suggested, for example, that a $d$-wave admixture in the nuclear wave function
can resolve this discrepancy, see \eg~Ref.~\cite{PhysRev.70.41}.
The wave function in LO \eftnopi~of the $A=2,3$ nuclei, however, 
has no $d$-wave component. Therefore, such explanations are excluded from our theory.
As we have shown, this limitation is compensated by the two-body currents, 
that reconcile the theory with the experimental/LQCD data.
\subsection{Magnetic polarizabilities}\label{sec:pol}
In general, polarizabilities parameterize the second-order response of a system to
an external probe. The dominant terms, which are quadratic in the magnetic field,
are provided in the \eftnopi~
formalism by an additional insertion of the one and two-body magnetic-moment
couplings as given in Eq.~\ref{eq:1b}~and~\ref{eq:2b}. The system is thereby subjected to the probe
at different points in space time, and the polarizability is then sensitive to its
deformation. In coordinate-space Schr\"odinger-equation practice, the calculation
is analogous to a second-order perturbation of the energy, see Appendix~\ref{sec.app2}. 
Again, the zero-range
approximation in the two-nucleon case allows for an analytic derivation of the
cutoff dependence of this quantity. This estimate was made in \cite{Drachman2016},
and yields a cutoff-independent polarizability of the deuteron.
 
\begin{figure}[h]
  \centering
  \begin{minipage}[t]{0.49\textwidth}
\includegraphics[width=1.0 \textwidth]{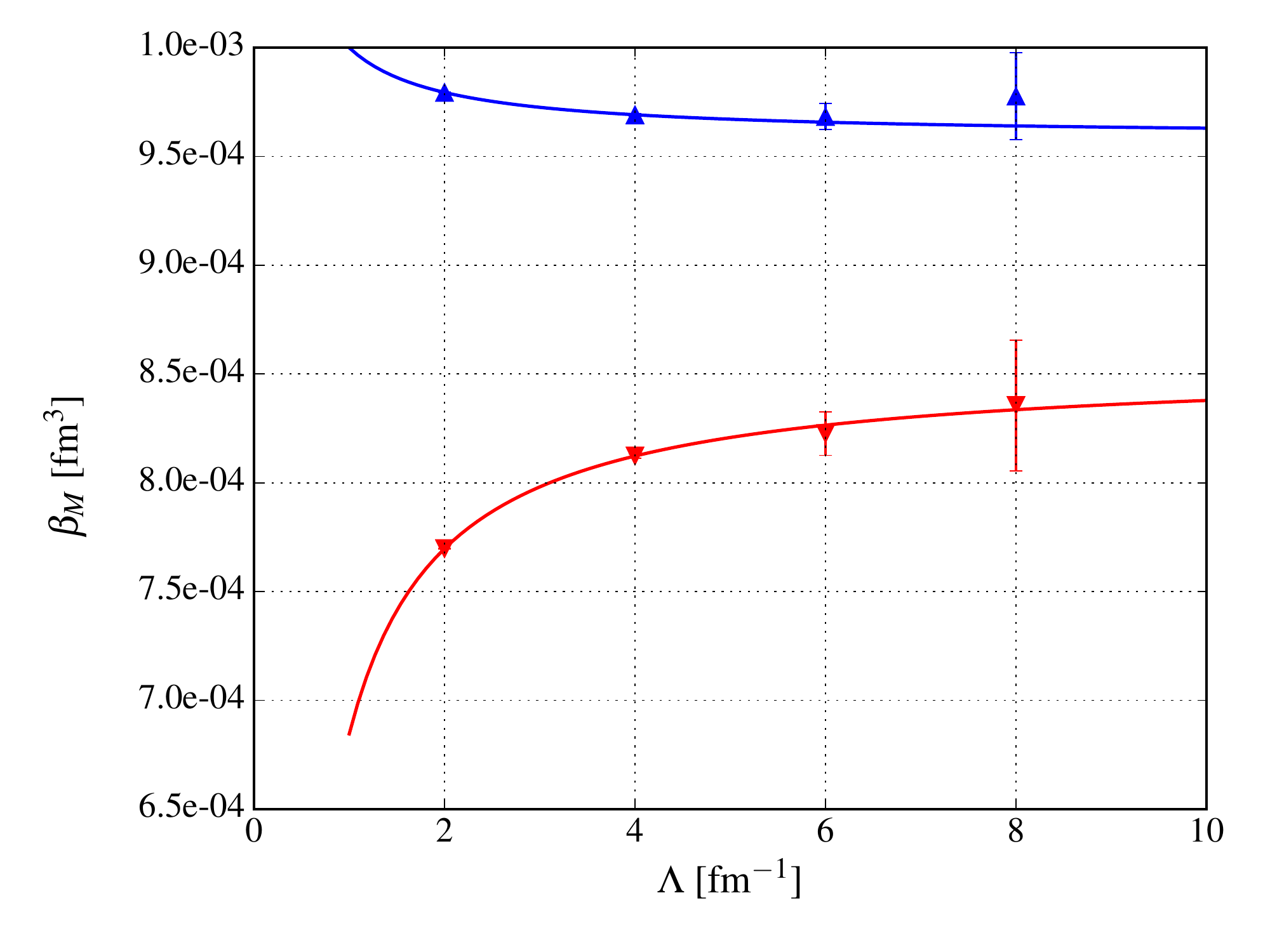}
\caption{\label{fig:pol140}(Color online) Regulator dependence of the magnetic polarizability
EFT calculations for $^3$He and triton  \mpis.}
  \end{minipage}
  \hfill
  \begin{minipage}[t]{0.49\textwidth}
\includegraphics[width=1.0 \textwidth]{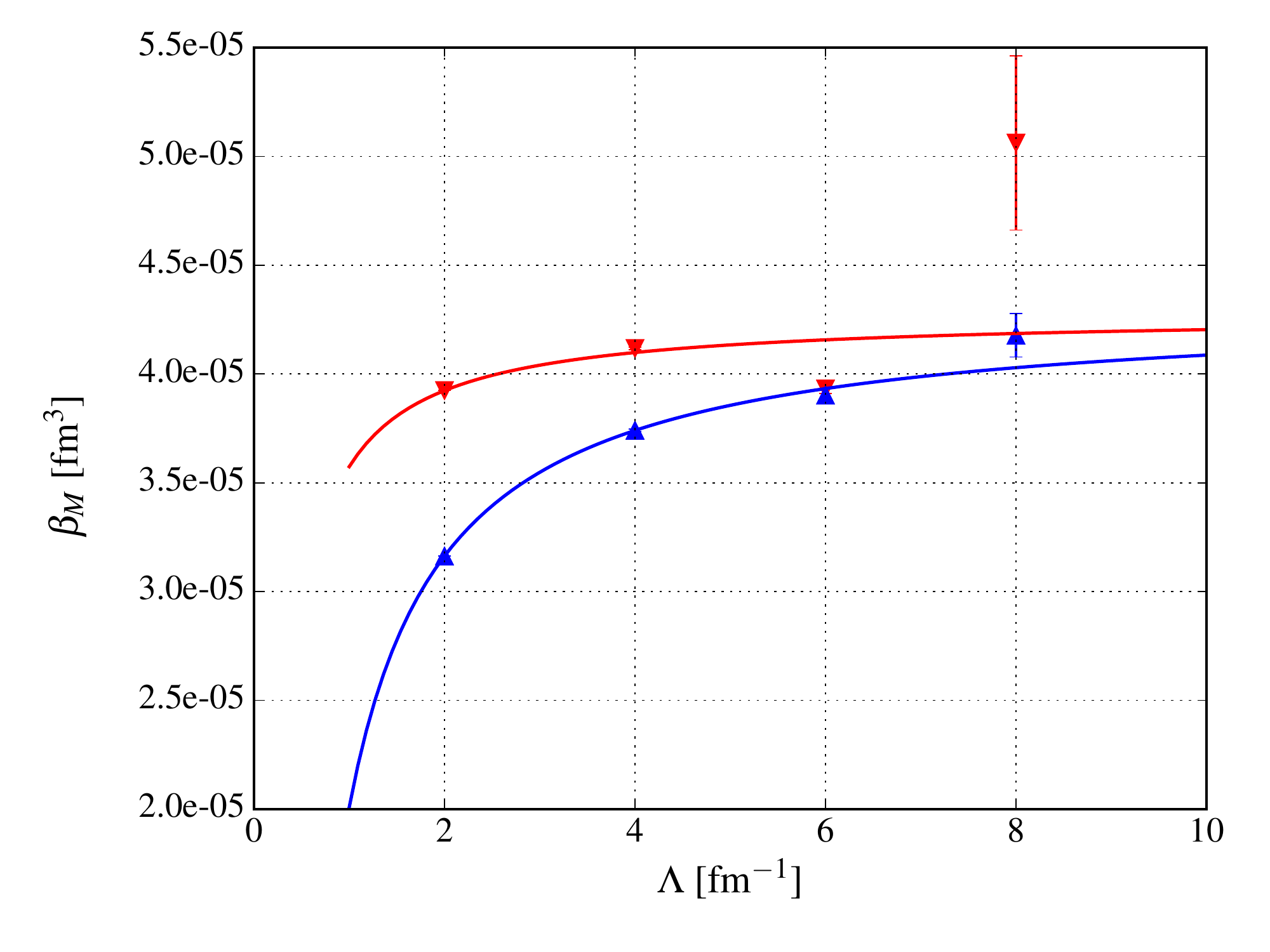}
\caption{\label{fig:pol860} (Color online) Regulator dependence of the magnetic polarizability
EFT calculations for $^3$He and triton  \mpil.}
  \end{minipage}
\end{figure}

The results for the magnetic polarizability of the deuteron $\pode$, triton $\pot$,
and helium $\poh$ are listed in Table~\ref{tab:pred}. 
In Figs.~\ref{fig:pol140}~and~\ref{fig:pol860}, we compare the regulator dependence of the 
polarization for the two $A=3$ mirror nuclei, $^3$He and triton at \mpis~and~\mpil.
The functional dependence
for interpolating the data points  was chosen as $a_1 +a_2/\Lambda^2$, where $a_1$ and $a_2$ are two 
constants employed to fit the data. 
The numerical accuracy, indicated by error bars in the figures, was used 
as a measure of the importance of the different data points in the fit. 

At \mpis, our postdictions for $\pode$ are consistent with previous
theoretical analyses and extractions based on cross-section data (see Table~\ref{tab:pred}).
The absolute value of $\pode$ is two orders of magnitude larger than the single-nucleon
polarizabilities and justifies, in part, why we call the deuteron a shallow nucleus.
Our predictions\footnote{To our knowledge, these numbers are first-time predictions and thus cannot
be compared with others.} for $\pot$~and $\poh$~signify relatively compact, rigid three-nucleon
bound states because they are of the same order of magnitude as $\beta_{n/p}$.

At \mpil, all polarizabilities, neutron/proton, deuteron (with $j_z=\pm1$), and the three-nucleon states,
are found by LQCD to be of the same order of magnitude. In particular, this entails
a deuteron which is by that measure as rigid, compact as the one and three-nucleon
states. This rigidity is consistent with the relatively large deuteron binding energy at \mpil.
The EFT postdictions, in turn, suggest a different response. For $j_z=\pm 1$ we get $\pode\approx 0$,
but for $j_z=0$ we find $\pode$ two orders
of magnitude larger than $\beta_{p/n}$ and therefore relatively pliant, like at \mpis.
Furthermore, we postdict $\pot$ and $\poh$ an order of magnitude smaller than the LQCD
predictions. Even the relatively large numerical uncertainty (see $\pot$ at $\Lambda=8$~fm$^{-1}$
in Fig.~\ref{fig:pol860}) cannot account for this difference.

\section{Summary}
\label{sec:summmary}
We have analyzed the pion-mass dependence of
magnetic moments, charge radii, and polarizabilities
of the deuteron, triton, and helion as characteristics of nuclei
in external electromagnetic fields.
The observables were calculated model-independently according to the pionless-effective-field-theory
formalism as developed for physical few-nucleon systems.
For unphysical pion masses, calculations were based on a previously applied match of this
theory to lattice QCD data. The robustness of the results with respect to different models
to account for the electromagnetic interaction within two-proton systems was assessed.

Results which pertain to physical nuclei are consistent with data and
previous calculations. The polarizabilities of the triton and helion are included as
predictions awaiting experimental verification.

For the analysis of lattice data at \mpim, we calculated the dependence of
the triton's magnetic moment on its binding energy. This dependence is found to
approach the shell-model limit at large binding energies and to decrease linearly up to a
discontinuity at the deuteron-neutron threshold. The relatively small slope of the
linear dependence leads to a prediction of the magnetic moment of the triton and helion.
A conjectured triton binding energy based on this prediction is found consistent with
a linear dependence of this energy on the pion mass.

Charge radii and magnetic moments of two-proton-nuclei are found
insensitive with respect to different models for the electromagnetic
interaction between constituent protons relative to the
accuracy which is expected from a NLO EFT analysis.
Nuclei at larger pion masses are found to be more robust in the two scenarios
we used to estimate the effect of dynamical QED.

In terms of the magnetic polarizability, we found the deuteron much more pliable
relative to the one and three-nucleon QCD calculations, and of the same order of magnitude
as the physical deuteron. 
%=============================================================================
\begin{acknowledgments}
We are very grateful for illuminating discussions with B.~C.~Tiburzi. 
  This work was supported by the Israel Science Foundation (Grant No.
  1308/16), the Pazi Research Foundation, and the NSF (Grant No. PHY15-15738).
\end{acknowledgments}
%=============================================================================
\appendix
%\renewcommand{\thechapter}{A}
%=============================================================================
\section{The low energy constants}
\label{sec:lecstbl}
In the following table we list the LECs used in our calculations.
\begin{table}[h]
\begin{center}
\caption{The 
%leading order 
LECs $c^\Lambda_{S,T,pp}$ , $d^\Lambda_3$ [GeV] and $l_{1,2}^\Lambda$ [n.d.] 
for physical ($m_{\pi}=140$ MeV) and lattice
($m_{\pi}=450,\;806\;\rm{MeV}$) nuclei 
%at \mbox{$m_{\pi}=140,\; 510,\;805\;\rm{MeV}$},
for various values 
%as function 
of the momentum cutoff 
\mbox{$\Lambda$ [fm$^{-1}$]}.}
\label{tbl:LECs}
\begin{tabular}
{c@{\hspace{5mm}} c@{\hspace{5mm}} c@{\hspace{5mm}} c@{\hspace{5mm}} c@{\hspace{5mm}} c@{\hspace{5mm}} c@{\hspace{5mm}} c}\hline\hline
$m_{\pi}$ & $\Lambda$ &  $c^\Lambda_T$  &  $c^\Lambda_S$ & $d^\Lambda_3$ & $c^\Lambda_{pp}$ & $l_1$ & $l_2$ \\
\hline
  140   &$2$ &$-0.1423$ &$-0.1063$ &$0.06849$ &$-0.0008303$ &$2.530$   &$-0.4652$ \\ 
        &$4$ &$-0.5051$ &$-0.4350$ &$0.6778$ &$-0.007646$ &$0.7349$  &$-0.1086$   \\ 
        &$6$ &$-1.091$ &$-0.9863$ &$2.653$ &$-0.01685$ &$0.3588$  &$-0.04717$ \\ 
        &$8$ &$-1.899$ &$-1.760$ &$7.816$ &$-0.02750$ &$0.2125$  &$-0.02617$  \\ 
        &$10$ &$-2.929$ &$-2.757$ &$20.48$ &$-0.03917$ &$0.1403$  &$-0.01660$  \\ 
        &$12$ &$-4.182$ &$-3.976$ &$50.94$ &$-0.05202$ &$0.09932$ &$-0.01152$  \\ 
        &$15$ &$-6.480$ &$-6.222$ &$195.6$ &$-0.07200$ &$0.06470$ &$-0.007324$ \\
\hline
  450   &$2$ &$-0.1637$ &$-0.1574$ &$  0.1580$ &$-0.003267$ &$2.023$ &$ 0.0288$ \\ 
        &$4$ &$-0.4837$ &$-0.4730$ &$  0.8374$ &$-0.009155$ &$0.556$ &$-0.00168$ \\ 
        &$6$ &$-0.9741$ &$-0.9591$ &$  2.711$ &$-0.01653$ &$0.269$ &$-0.00207$ \\ 
        &$8$ &$-1.635$  &$-1.616$  &$  7.182$ &$-0.02494$ &$0.160$ &$-0.00150$ \\ 
        &$10$ &$-2.466$  &$-2.443$  &$ 17.33$ &$-0.03422$ &$0.106$ &$-0.00107$ \\ 
        &$12$ &$-3.468$  &$-3.440$  &$ 40.04$ &$-0.04421$ &$0.075$ &$-0.000843$ \\ 
        &$15$ &$-5.291$  &$-5.256$  &$137.0$ &$-0.06032$ &$0.049$ &$-0.000579$ \\
\hline
  806   &$2$  &$-0.1480$ &$-0.1382$ &$ 0.07102$&$-0.002125$ &$1.476$ &$0.5907$\\
        &$4$  &$-0.4046$ &$-0.3885$ &$ 0.3539$ &$-0.006886$ &$0.3017$ &$0.1199$\\
        &$6$  &$-0.7892$ &$-0.7668$ &$ 1.001$  &$-0.01298$  &$0.1242$ &$0.0492$\\
        &$8$  &$-1.302$  &$-1.273$  &$ 2.221$  &$-0.02007$  &$0.06710$ &$0.02656$\\
        &$10$ &$-1.942$  &$-1.907$  &$ 4.308$  &$-0.02814$  &$0.04194$ &$0.01660$\\
        &$12$ &$-2.710$  &$-2.670$  &$ 7.712$  &$-0.03676$  &$0.02860$ &$0.01130$\\
        &$15$ &$-4.103$  &$-4.052$  &$16.84$   &$-0.05077$  &$0.01805$ &$0.007092$\\
%%% L1 fitted to mu(3H)
%%%  806   &$2$  &$-0.1480$ &$-0.1382$ &$ 0.07102$&$-0.002125$ &$1.447$ &$0.5907$\\
%%%        &$4$  &$-0.4046$ &$-0.3885$ &$ 0.3539$ &$-0.006886$ &$0.3530$ &$0.1199$\\
%%%        &$6$  &$-0.7892$ &$-0.7668$ &$ 1.001$  &$-0.01298$  &$0.1667$ &$0.0492$\\
%%%        &$8$  &$-1.302$  &$-1.273$  &$ 2.221$  &$-0.02007$  &$0.09932$ &$0.02656$\\
%%%        &$10$ &$-1.942$  &$-1.907$  &$ 4.308$  &$-0.02814$  &$0.06631$ &$0.01660$\\
%%%        &$12$ &$-2.710$  &$-2.670$  &$ 7.712$  &$-0.03676$  &$0.04756$ &$0.01130$\\
%%%        &$15$ &$-4.103$  &$-4.052$  &$16.84$   &$-0.05077$  &$0.03145$ &$0.007092$\\
\hline\hline
\end{tabular}
\end{center}
\end{table}

%=============================================================================
\section{Magnetic moments in the zero-range limit}
\label{sec.app1}
The analysis of the two-nucleon system based on an interaction constrained by a single
datum, namely the deuteron binding energy, was instigated almost a century ago in
Ref.~\cite{10.2307/96363}. What later became known as the zero-range
approximation can be used here to derive analytically the dependence of the two-body-current
LECs $l_{1},l_2$ as introduced in Eqs.~\ref{eq:lagr},\ref{eq:2b}.

The bound-state solution of the Schr\"odinger equation in an area of vanishing potential
reads
\ber
\label{eq:wf}
\bra r~|~\text{BS}~\ket&=& {A_S \over \sqrt{4 \pi}} {e^{-\kappa r} \over r}\;\;,
\eer
where $A_s$ is the wave function normalization and $\kappa=\sqrt{m B }$ is set by the deuteron's
 (dineuteron's) binding energy $B_D$ ($B_{nn})$.

The contribution of the one-body current as parameterized in Eq.~(\ref{eq:1b}) is evaluated
to be
\be
\label{eq:re1bm}
\bra\text{BS}~|~\bs{\mu}^{(1)}~|~\text{BS}~\ket={A_S^2 \over 2 \kappa}\mu_N(g_p+g_n)\;\;.
\ee
Similarly, the two-body current regularized with a Gaussian,
Eq.~(\ref{eq:2b}), yields the following result for the spin-triplet state 
\be\label{eq:2bd}
  \bra\text{BS}~|~\bs{\mu}^{(2)}~|~\text{BS}~\ket={A_s}^2 \mu_N l_2 \Lambda^2.
\ee
Cutoff independence implies $l_2 \propto \Lambda^{-2}$. This regulator dependence
was found above (see discussion of Fig.~\ref{fig:cstpp}) numerically.
We can compare these expressions with the \eftnopi~ calculation of \cite{Chen:1999tn}
where the authors used a power-divergence-subtraction method introducing a
dimensional regularization scale $\mu$,
\be\label{eq:reschen}
  \mu_D=\mu_N(g_p+g_n)+\tilde{l}_2\sqrt{m B_D}\left(\mu-\sqrt{m B_D}\right)^2.
\ee
These results coincide in the zero-range limit where
in which the asymptotic wave function is normalized to 1, and
$A_S^2\longrightarrow 2\sqrt{mB_D}$.
The $\mu$ dependence of the NLO LEC can be determined for arbitrary values
of $\mu$ but will coincide with the $\Lambda$ dependence for $\mu\gtrsim m_\pi$.
\section{Magnetic polarizabilities}
\label{sec.app2}
The calculation of polarizabilities as parameterizations of the second-order 
response of a nucleus (spin-quantum numbers $j_0,~m_0$) to perturbation given
by its coupling to an external magnetic field is explained here.
Specifically, the twice-iterated coupling of the photon to the nucleus
shifts its energy by an excitation of intermediate states $n$:
\ber
\Delta E^{(2)}&=&\sumint_n\frac{\bra j_0m_0~\vert~\ve{\mu}\cdot\ve{B}~\vert~ j_nm_n~\ket\bra j_nm_n~\vert~\ve{\mu}\cdot\ve{B}~\vert~ j_0m_0~\ket}{E_n-E_0}\;\;\nonumber\\
&\equiv&\frac{1}{2}\sum_{\lambda\nu}(-)^\nu\beta^{(\lambda)}_\nu~B^{(\lambda)}_\nu\;\;.
\eer
Thereby, the spherical components of the polarizability
\be
\beta_\nu^{(\lambda)}=\frac{2}{3}\sumint_n\frac{\vert\bra j_0~\vert\vert~\ve{\mu}~\vert\vert~ j_n~\ket\vert^2}{E_n-E_0}\sum_q(-)^q\bra j_0m_0j_nm_n~\vert~1~q~\ket^2\bra 1q1-q~\vert~\lambda\nu~\ket\;\;,
\ee
and the quadratic field tensor
\be
B_\nu^{(\lambda)}=(-)^\nu\sum_{pq}\bra 1p1q~\vert~\lambda\nu~\ket~ B_pB_q
\ee
are defined. For $\ve{B}=B\ve{e}_z$, the expression of the shift in terms of scalar and
tensor polarizability is
\be
\Delta E^{(2)}=\left(-\frac{1}{2\sqrt{3}}\beta_0^{(0)}+\frac{1}{\sqrt{6}}\beta_0^{(2)}\right)~B^2
\ee
with
\ber
  \beta_0^{(0)}&=& -\frac{2}{3}\frac{\sqrt{3}}{{2j_0+1}}
                  \sumint_n\frac{\vert\bra j_0~\vert\vert~\ve{\mu}~\vert\vert~ j_n~\ket\vert^2}{E_n-E_0}
\cr
  \beta_0^{(2)}&=&
       -12\sqrt{5}\frac{m_0^2-\frac{1}{3}j_0(j_0+1)}{\sqrt{(2j_0+3)(2j_0+2)\dots(2j_0-1)}}
       \sumint_n\frac{\vert\bra j_0~\vert\vert~\ve{\mu}~\vert\vert~ j_n~\ket\vert^2}{E_n-E_0}
       \textsf{W}(j_nj_012;1j_0)\;\;.
%%% \cr
%%%   \beta^{(2)}_0 &=& 
%%%             12\sqrt{5}\frac{m_0^2-\frac{1}{3}j_0(j_0+1)}
%%%             {\sqrt{(2j_0+3)(2j_0+2)\cdots (2j_0-1)}}
%%%             \sumint_n
%%%           (-)^{j_0+jn}
%%%           \sixj{j_n}{j_0}{1}{2}{1}{j_0}
%%%            \frac{|\bra j_0 || \bs{\mu} || j_n \ket|^2}{E_n-E_0}\;\;.
\eer
Weighted with Racah's \textsf{W}-coefficient, we combine matrix elements for the allowed transition
where care has to be taken to include the additional $j_n=0$ bound states at the unphysical $m_\pi$.
The definition of scalar and tensor polarizabilities is then identical to that used in
Ref.~\cite{Chang:2015qxa}.

%The matrix elements can be evaluated in the zero-range approximation with Eq.~\ref{eq:wf} for bound
%states and $\bra r~|~\text{SC}~\ket=(kr)^{-1}\sin(kr)-a(1+ika)^{-1}e^{ikr}/r$ for scattering states
%in a channel with scattering length $a$ and relative momentum $k$.
%The one-body operator induces transitions to singlet and triplet states. It does not
%couple triplet bound and triplet scattering states because of their orthogonality and
%character as eigenvectors of $\hat{V}^{(1)}_{\text{nucl}-B}$.
%\be
%\label{eq:deut_mag}
%\beta_d= {\alpha A_s^2\mu_N^2 (g_p-g_n)^2 (|y_s|+1/3)(y_s-1)^2 \over  4 \kappa_{d}^3 m (1+|y_s|)^3},
%\ee
%where $y_s\equiv \sqrt{mB_d} a_s$ and $a_s$ is the singlet scattering length.
%For $m_0=\pm1$, this element is only relevant if there is at least one more bound triplet state
%in the spectrum, otherwise, its various contributions add up to zero.
%At NLO, 
%\be 
%\label{eq:2bpold}
%\beta_{d}^{0 \rightarrow 1}= { \alpha M_N  l_2^2 A_s^2 \Lambda^4\over3 \pi } \left [{a_s  (%y_s^4-1)+1-3y_s^2+2y_s^3 \over 4 (y_s^2-1)^2} \right ]\;\;.%
%\ee%
%This result suggests a relatively weak $\Lambda$ dependence of the polarization%
%since the $\Lambda^{4}$ in Eqs. (\ref{eq:2bpold}) is canceled by the $\Lambda^{-2}$ dependence of 
%both the LECs $l_1,l_2$.
% -----------------------------------------------------------------------------
%=============================================================================
\bibliographystyle{hieeetr}
\bibliography{refs}
\end{document}